\documentstyle{memsait}
\input epsf.sty
\begin{opening}
\title{Infrared instrumentation for large telescopes: an alternative approach}
\author{E. Oliva$^1$}
\institute{$^1$Osservatorio Astrofisico di Arcetri, Largo E. Fermi 5,
I-50125 Firenze, Italy}
\date{} 
\end{opening}

\begin{document}

\oddpagefooter{}{}{} 
\evenpagefooter{}{}{} 
\ 
\bigskip

\begin{abstract}
I very briefly describe
the latest generation near infrared (1--2.5 $\mu$m) instruments
which are available on, or under development for `large' ($D\!\ge\!3.5$ m)
telescopes.\\
Most of the imagers under construction are limited to
relatively small fields, while the spectrometers  aim
at quite high resolving powers.
The alternative instruments which I discuss here are\\
-- WIDE, a relatively low--cost instrument for the prime focus of LBT
and/or of TNG optimized
for deep imaging of very large fields (12'$\times$12' on LBT and 
26'$\times$26' on TNG) through
the 1~$\mu$m, J, H, K' broad--band filters.\\
-- AMICI, an ultra--high efficiency, low resolution disperser 
optimized for collecting complete 0.9-2.5 $\mu$m  spectra of very
faint objects. This device is mounted in NICS (the IR instrument for TNG) and 
should soon deliver spectra with quality comparable to that obtained 
with instruments on 8m class telescopes with similar integration times.

\end{abstract}

\section{Infrared imagers}

\begin{table}[!ht]
\hspace{0cm}
\newcommand{\X}{\mbox{$\;\times\;$}}
\centerline{\bf Table 1 -- IR Imagers working or planned on large telescopes}
\vskip10pt
\begin{flushleft}
\begin{tabular}{|l|l|c|r|r|l|}
\hline
& & & & & \\ 
Telescope$^{a}$  &  Instrument$^{b}$  &  
      f.o.v.$^{c}$  & Speed$^{d}$ & Ref.$^{e}$   &  Comments \\
& & & & & \\ 
\hline

& & & & & \\ 
AAO       &    IRIS2       & 7.7'\X7.7' & 3.1 & T1 &  \\

WHT       &    CIRSI       & 11'\X11' & $\simeq$3$^f$ & T2 
   & Operating, $\lambda\!<\!1.8$  $\mu$m,\\
WHT       &    INGRID      & 4.2'\X4.2'& 0.92 & T3 &  \\

KPNO   &    IRIM        &  2.5'\X2.5' & 0.33 &  T4 & Operating  \\

CTIO   &   OSIRIS       &  3.9'\X3.9' & 0.80 & T5 & Operating \\
CTIO   &   CIRIM        &  1.7'\X1.7' & 0.15 & T6 & Operating \\

CFHT      &    KIR         & 0.6'\X0.6' & 0.015 & T7 & Operating \\
CFHT      &    RedEye      &  2'\X2'    & 0.16 & T8 & Operating \\

Calar-Alto &   $\Omega$-prime &  6'\X6'  & 1.4 & T9 &  Operating? \\
Calar-Alto &   $\Omega$-Cass  &  5'\X5'  & 1.0 & T10 &  Operating \\

UKIRT     &    IRCAM3   &   1.2'\X1.2' & 0.07 & T11 & Operating \\
UKIRT     &    UFTI     &   1.6'\X1.6' & 0.13 & T12 & Operating \\
                       
NTT       &    SOFI     &   5'\X5' & 1.0 & T13 & Operating   \\
TNG       &    NICS      & 4.2'\X4.2' & 0.71 & T14 & \\

Palomar & P.F. IR cam. & 2'\X2' & 0.33 & T15 & Operating (private instr.) \\
Palomar & Cass. IR cam. & 0.6'\X0.6' & 0.03 & T16 & 
                 Operating (private instr.) \\

MMT    &  CIRSI     &   5.1'\X5.1'  & 3.1 & T17 & Proposed in 1996, \\
& & & & & cancelled? \\
MMT    &  CfA IR cam. & 6.8'\X6.8' & 5.4 & T18 & Concept design, \\
& & & & & phase A started? \\

VLT-UT1   &    ISAAC      &  2.5'\X2.5' & 1.4  & T19 &  Operating \\
VLT-UT4   &    NIRMOS     &  12'\X16'   & 43 & T20 & 
  Limited to $\lambda\!<\!1.8$ $\mu$m \\

Keck1/2   &      NIRC1/2   &   0.64'\X0.64' & 0.13 & T21 & Operating  \\

Gemini    &    NIRI       &   2'\X2'  & 0.88  & T22 & \\

Subaru    &    IRCS       &   1'\X1' & 0.22  & T23 & \\

LBT       &    LUCIFER    &   4'\X4'   & 3.7 &  & Phase A started \\

& & & & & \\  
LBT       &    WIDE--LBT  &  12'\X12'  & 33 & T24 & 
                             Proposed to CNAA, \\
TNG       &    WIDE--TNG  &  26'\X26'  & 27 &  &  phase A completed \\

& & & & & \\  

\hline
\end{tabular}
\vskip3pt
\def\NOTA#1#2
{\hbox{\vtop{\hbox{\hsize=0.1cm\vtop{#1}}}
      \vtop{\hbox{\hsize=13cm\vtop{#2}}}}\vskip3pt}

\vskip5pt
\NOTA{$^a$}{ Telescopes with $D\!\ge\!3.5$ m which are operating
or expected to work in the next few years.}

\NOTA{$^b$}{ Name of the IR imager available or planned witihin the next 
few years }

\NOTA{$^c$}{ Field of view on sky, in arc-minutes.}

\NOTA{$^d$}{ Speed factor, i.e. time needed to image a given f.o.v. to
a given depth, normalized to NTT-SOFI}

\NOTA{$^e$}{ See the reference list }

\NOTA{$^f$}{ CIRSI is an "only detector camera" which works  
with pre--existing optical correctors that have low transmissions
in H, typically a factor of $\sim$2 lower than IR optimized lenses. }

\end{flushleft}
\end{table}

Deep multicolour imaging of large fields is one of the most important
and popular tools for studying a large variety of astrophysical objects.
This simple recognition, together with the recent availability of
very large format CCD detectors, has prompted many groups to develop wide 
field optical cameras for large telescopes. An excellent analysis 
of the status and performances of these instruments can be found in the
WFI--LBT report (Giallongo et al. 1999) which also contains an exhaustive
discussion of the scientific cases for deep wide field imaging.\\
While the astronomical community is taking (or will soon take) advantage from 
several powerful wide field optical imagers,
the situation for imaging in the near infrared (1--2.5 $\mu$m) is much
less encouraging. Table 1 is a list of all the NIR instruments working,
or planned for the next $\simeq$5 years on large telescopes.\\
 In spite of the fact that
the latest generation 1024$^2$ (and soon 2048$^2$) IR arrays allow
coverage of about 10'$\times$10' fields at seeing--limited
resolutions, the average field covered by the NIR instruments is much lower.
In particular, the situation on the largest telescopes is far from encouraging.
On the upgraded MMT, a 5'$\times$5' camera was proposed 
by the Cambridge's
group in 1996 but, to the best of my knowledge, this project has
been cancelled. A similar instrument is now proposed by the CfA 
but its status is still very unclear (see ref. T18).\\
On 8--10m class telescopes 
no wide field imager is officially planned 
apart from NIRMOS, which is indeed a multi--object spectrometer and will 
be seldom used as an imager. \\
The main consequence of such a situation is that all deep imaging surveys
will be severely biased toward sources which are blue enough to be detected
by CCDs, while miss intrinsically red objects such as very cool 
brown dwarfs, elliptical galaxies at $z\!>\!1.5$ and QSOs at $z\!>\!10$
(just to mention a few of the ``hottest'' subjects).
This problem could be alleviated if an instrument like WIDE 
becomes operative on either the LBT and/or the TNG telescopes.
In both cases, the ``survey power'' would be
a factor $\ge$10 larger than any other instrument available or planned
(see Table 1).
I report here the main results of the phase--A study of the
WIDE instrument.\\

\begin{table}[!t]
\hspace{0cm}
\centerline{\bf Table 2 -- Cost estimates for WIDE}
\vskip10pt
\begin{flushleft}
\begin{tabular}{|l|rr|l|}
\hline
 & & & \\
Item & \multicolumn{2}{c|}{cost$^a$} & Firm\\
     & LBT & TNG & \\
 & & & \\
\hline
 & & & \\
Glass optics and lens holders & 215 & 175
     & Gestione SILO (Firenze) \\
 & & & \\
Crystal optics & 190 & 175 & Janos Techn. (USA) \\
 & & & \\
Crystotat and mechanics & 190 & 220 & Rial (Parma) \\
 & & & \\
Filters (1 $\mu$m, J, H, K') $\oslash$80mm  &  45 & 120 & Barr Ass. (USA) \\
 & & & \\
Electronics--software & 130 & 130 & various\\
 & & & \\
Total (w/o array) & 770 & 820 & \\
 & & & \\
\hline
 & & & \\
Array & $\simeq$600$^c$ & $\simeq$1500 & Rockwell \\
 & & & \\
\hline
\end{tabular}
\vskip5pt

\def\NOTA#1#2
{\hbox{\vtop{\hbox{\hsize=0.015\hsize\vtop{#1}}}
      \vtop{\hbox{\hsize=0.97\hsize\vtop{#2}}}}\vskip3pt}
\NOTA{$^a$}{ In millions of Italian Lire}
\NOTA{$^b$}{ Cost of single array could be significantly reduced 
if purchase order
is coordinated with WIDE--TNG or with groups working on other instruments
(e.g. Lucifer, Oneiric)}
\end{flushleft}
\end{table}

\section{The WIDE instrument}

\begin{figure}
\epsfysize=6.8cm
\hspace{1.0cm}\epsfbox{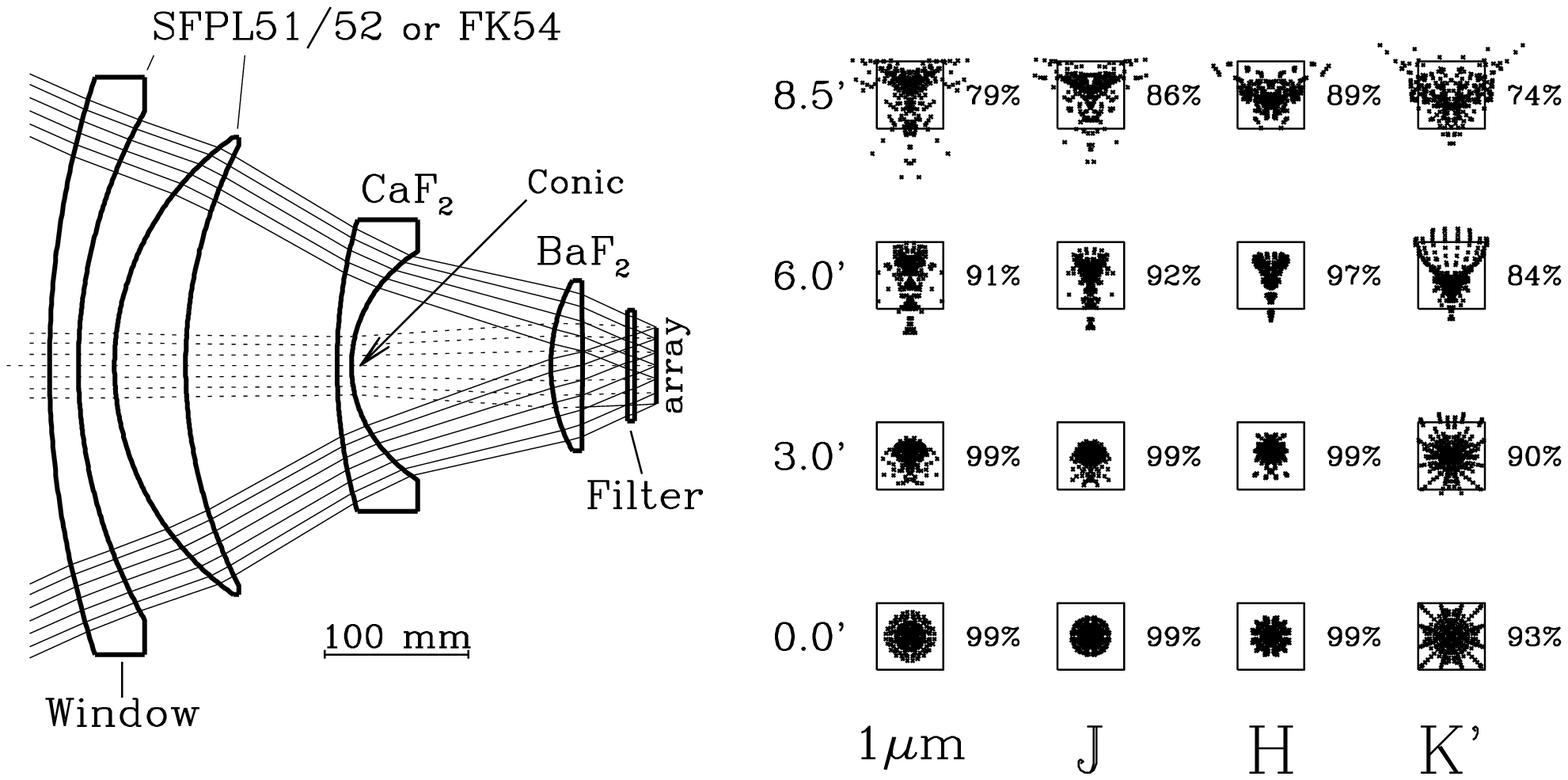}
\caption[h]{ 
Left: optical layout of {\bf WIDE--LBT}, the 12'$\times$12' 
IR camera for the prime focus of the LBT telescope. The first 
$\oslash$400 mm lens
has very lax centering/tilt tolerances and acts also as window
for the dewar.\hfil\break
Right: Polychromatic spot diagrams for imaging through the
1 $\mu$m (0.95--1.1 $\mu$m), J (1.1--1.4 $\mu$m), H (1.5--1.8 $\mu$m)
and K' (1.95--2.3 $\mu$m) filters.
The squares are 18.5$\times$18.5 $\mu$m (equivalent to 0.352"$\times$0.352"
on the sky) and correspond to the size of 1 pixel
of the Rockwell HgCdTe array.
The spots are shown at various positions from the center (0') to the corner
(8.5') of the array and the numbers are the fraction of energy falling
within a circle of $\oslash$18.5 $\mu$m.
The distortion is 0.5\% at 6' (array edge) and 1.0\% at 8.5' (field corners).
}
\end{figure}

The main goal of the WIDE project
is to build a simple and relatively inexpensive NIR instrument which could
cover the largest possible field of view for seeing limited
imaging on 3.5m and  8m class telescopes.\\
The prime focus of the LBT, with a natural scale of 21"/mm (i.e.
0.38"/pix on a Rockwell HgCdTe array), is the ideal site for such 
an instrument. Simple considerations on the
relative roles of airglow and thermal backgrounds, together with
the recent experience of the $\Omega$--prime 
instrument at Calar Alto, indicate that
the prime focus is an excellent station to perform IR imaging 
in the airglow dominated bands, i.e.  from 1~$\mu$m  to K'.
Moreover, a prime focus camera is much simpler and consists of much fewer 
optical elements than Cassegrain instruments with a similar field of view.\\
More details on the expected performances can be found in the original
WIDE proposal that was submitted to the CNAA in May 1998 (see ref. T24)
 which also includes a quite detailed analysis of the various 
technological aspects of this instrument.
In the last year we concentrated on the opto--mechanical design and 
verified the feasibility (and estimated the cost) of the various
parts by contacting various companies.\\
Another excellent possibility is to exploit the prime focus of the TNG
telescope, in which case it is necessary to use a mosaic of 4x4 arrays
to cover an area large enough to achieve a survey power similar to the LBT.
Figs.~1,2 show the optical layouts of the instruments. The larger lenses 
(max $\oslash$ 400 mm) are
manufactured out of standard fused silica (IR grade) or glasses from
the Ohara Corp. (SFPL51 and
SPFL52) or Schott (FK54) catalogues. 
All these glasses have negligible internal absorptions
at $\lambda\!<\!2.4$ $\mu$m.\\
%

\begin{figure}
\epsfysize=6.8cm
\hspace{1.0cm}\epsfbox{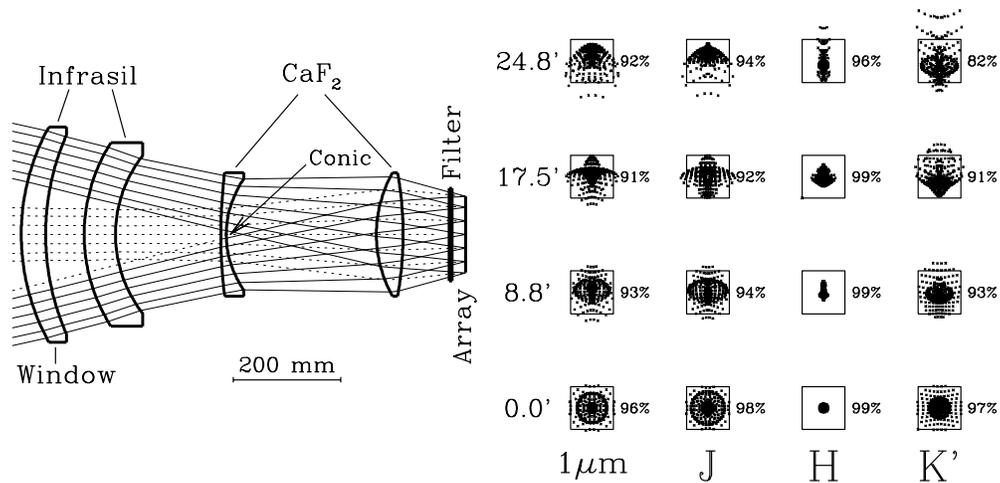}
\caption[h]{ 
Left: optical layout of {\bf WIDE--TNG}, the 4$\times$13'$\times$13' 
IR camera for the prime focus of the TNG telescope. 
The focal length is 10~m which yields a scale of 0.382"/pix on
a Rockwell HgCdTe array.
The total corrected field of view is 35'$\times$35' and can accomodate
a mosaic of four separated (and losely spaced) 2048$^2$ detectors.
The ``Filter'' is a mosaic of four 45$\times$45 mm elements, i.e. with
sizes which are well within the capabilities of filter manufacturers.
\hfil\break
Right: Polychromatic spot diagrams for imaging through the
1 $\mu$m (0.95--1.1 $\mu$m), J (1.1--1.4 $\mu$m), H (1.5--1.8 $\mu$m)
and K' (1.95--2.3 $\mu$m) filters.
The squares are 18.5$\times$18.5 $\mu$m (equivalent to 0.382"$\times$0.382"
on the sky) and correspond to the size of 1 pixel
of the Rockwell HgCdTe array.
The spots are shown at various positions from the center (0') to the corner
(17.5') of the array and the numbers are the fraction of energy falling
within a circle of $\oslash$18.5 $\mu$m.
The distortion is 0.5\% at 17.5' (array edge) and 1.0\% at 24.8' 
(field corner).
}
\end{figure}

The smaller lenses are made of calcium or barium fluoride crystals which are
regularly produced in large blanks by several companies around the world. The
sizes of these lenses is quite standard. In particular, the
BaF$_2$ lens is slightly smaller than the collimator of ISAAC while the
CaF$_2$ elements are all significantly smaller than the lenses 
normally used in UV micro--lithography instruments. \\
The only non--spherical element is the first CaF$_2$ lens which has
a conical surface: K=--0.22 and K=--0.36 for LBT and TNG, respectively.
The sizes and sphapes are within the capabilities
of companies specialized in single point diamond machining 
(e.g. Janos Technology). However, it should be noted that the aspheric
on the TNG design, with a maximum deviation from sphere of 290 $\mu$m,
is much less demanding than that for LBT which deviates up to almost 
800 $\mu$m.\\
The system for LBT is virtually free from chromatism and the
image quality is excellent,
i.e.  $>$80\% of the light within one pixel,
over most of 12'$\times$12'
field of view covered by a single 2048$^2$ array (see Fig. 1). 
The image distortion is also quite good: 0.5\%
at the field edge (6' from axis) and 1.0\% at the corners (8.5' from
axis). \\
The design for TNG, which employs the much more dispersive (but cheaper)
infrasil glass, requires refocussing in the various bands and
provides excellent images (see Fig.~2) over a spectacularly large field
of view, namely 35'$\times$35', with an image distortion of only 0.5\% and
1\% at the field edges and corners, respectively.
This is sufficient to accomodate a mosaic of four
non--buttable 2024$^2$ array each of them covering an area of 13'$\times$13'.\\

A specific advantage of both designs is that the positioning of the
first lens has very lax 
tolerances: a decenter of 0.5 mm and/or a tilt of 0.1 degrees can
be fully compensated by shifting/tilting the whole dewar and, in practice,
 produce
a negligible effect on the image quality. Therefore, the first
lens can also act as the dewar window without requiring any special
mechanical mount. The deformations induced by the pressure difference
between the outside environment (air) and the inner vacuum amount
to several microns (cf. Fig. 3) but have a totally negligible effect
on the image quality.\\

\begin{figure}
\epsfysize=6.3cm
\hspace{2cm}\epsfbox{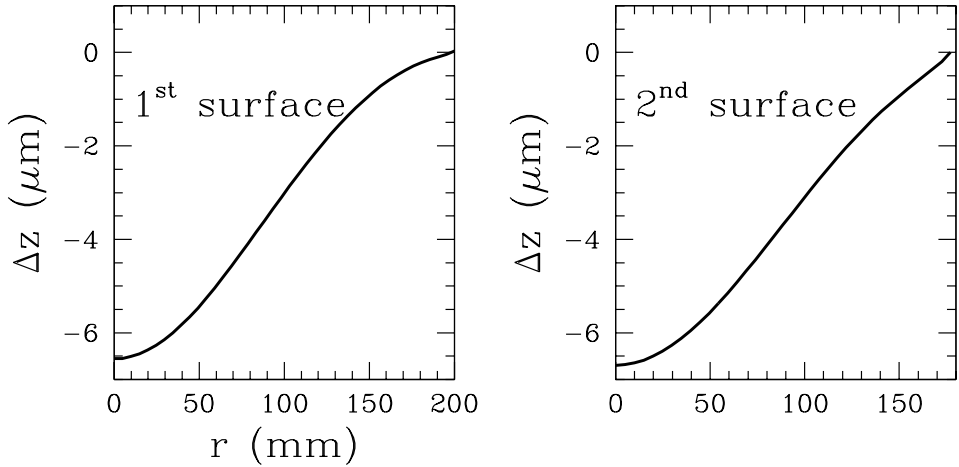}
\caption[h]{ Deformation of the first lens (see Figs.~1,2) when this
optical element is used as window of the dewar. The curves are based
on finite element analysis and include the effect of pressure difference
(1 atmosphere between the outside environment and the vacuum  tank)
and weight. The latter amounts to only $<$0.06 $\mu$m and is therefore
totally negligible.
}
\end{figure}

The cost estimate is summarized in Table 2 which also includes the names
of the companies which we already contacted for the various items.
Note that the overall cost of the instrument is dominated by the price
2048$^2$ Rockwell array(s), expecially in the case of the instrument for
the TNG which requires four such devices.

\begin{figure}
\epsfysize=6.0cm
\hspace{0cm}\epsfbox{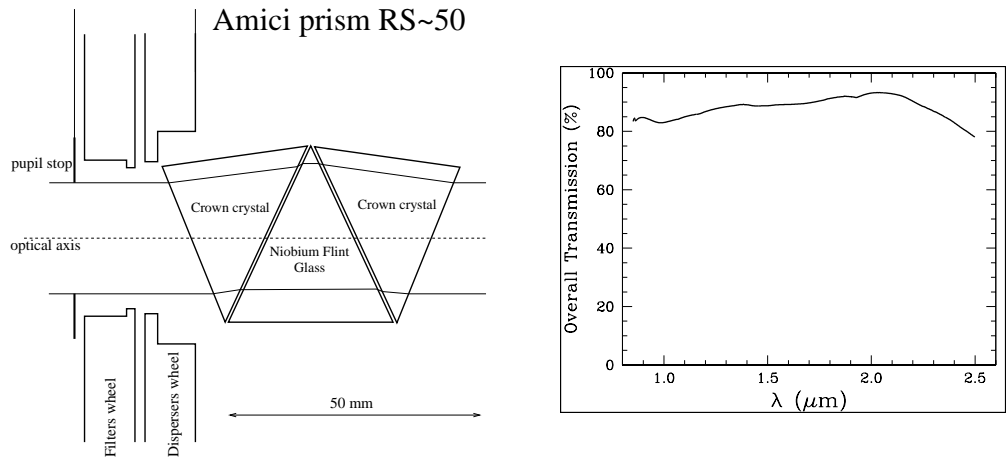}
\caption[h]{ 
Left: sketch of the AMICI disperser which consists of 3 prisms organized in the
classical Amici mount. The highly dispervive Flint prism, with a vertex angle
of 50$^o$, yields a value of RS (i.e. resolving power with a 1" slit) of
about 50. \\
Right: measured efficiency of the device, note that
the overall transmission exceeds 80\% over the entire 
0.85--2.45 $\mu$m wavelength range.
}
\end{figure}

\section{ Low dispersion spectroscopy: the AMICI device}

Low dispersion IR spectroscopy covering the widest possible wavelength
range is a fundamental tool for studying very faint objects with broad 
spectral features. These include:\\
-- elliptical galaxies at $z\!>\!1.5$ which can be recognized by the
4000 \AA\  break characteristic of relatively old stellar populations
(e.g. Soifer et al. 1999)\\
-- methane dwarf stars, i.e. brown dwarfs cooler than 1500 K and whose 
spectrum is characterized by the prominent CH$_4$ band--head at 1.6 $\mu$m as
well as by the very broad H$_2$O bands which extend into the J, H and K
bands (e.g. D'antona et al. 1999).\\
All the existing/planned IR spectrometers for large telescopes employ
gratings and/or grisms as light dispersers. This choice intrinsically limits
the spectral coverage to 1 or at most 2 photometric bands (e.g. J+H or H+K)
per frame. The average efficiency of the grating/grism 
over the spectral free range
is $<$50\%
including the losses introduced by the order sorter filter. \\
 
The alternative approach which we adopted in NICS, the IR instrument for
the TNG, is to use a prism--based disperser which is sketched in Fig.~4.
The Crown--Flint--Crown symmetrical combination corresponds to the classical
Amici mount with, however, separated (not glued) elements.
The resolving power is RS$\simeq$50 
and the average efficiency we achieved with ad--hoc multi--layer A/R
coatings exceeds 80\% (with peaks $>$90\%) over the full 
0.85--2.45 $\mu$m range (See Fig.~4).\\
 
To estimate the ``speed'' of this device it is convenient to compare the
NICS--AMICI combination with the ISAAC--LR spectroscopic mode. The latter
uses a grating disperser with average efficiency (within each band) of 50\%
and requires 4 different exposures to cover the 0.9--2.5 $\mu$m range.
The AMICI disperser has an efficiency a factor 1.8 higher and delivers
the full spectrum in a single shot. Therefore, the factor of $\simeq$7
gain in time one has using AMICI on the TNG should fully compensate the factor
of 5 loss due to the lower area of the TNG relative to the VLT. In other
words, AMICI on the TNG should soon produce low resolution spectra 
with similar quality, and with similar integration times,
as ISAAC on the VLT.

\end{document}